\documentstyle[prb,aps,epsfig]{revtex}

\begin{document}
\draft

\title{Metastable states in ultrathin magnetic films} 

\author{Pablo M. Gleiser, Francisco A. Tamarit\cite{auth2} and Sergio A. Cannas\cite{auth2}}

\address{Facultad de Matem\'atica, Astronom\'\i a y F\'\i sica, 
         Universidad Nacional de C\'ordoba, Ciudad Universitaria, 5000 
         C\'ordoba, Argentina}
\date{\today}
\maketitle

\begin{abstract}
 The equilibrium phase diagram of a two-dimensional Ising model with competing exchange and dipolar interactions is analyzed using a Monte Carlo simulation technique. We consider the low temperature region of the $(\delta,T)$ phase diagram ($\delta$ being the ratio between the strengths of the exchange and dipolar interactions) for the range of values of $\delta$ where striped phases with widths $h=1$ and $h=2$ are present. We show that the transition line between both phases is a first order one. We also show that, associated with the first order phase transition, there appear metastable states of the phase $h=2$ in the region where the phase $h=1$ is the thermodynamically stable one and {\it viceversa}.
\end{abstract}


The physics of ultrathin  films and quasi-two-dimensional systems is of current interest because of its many technological applications. For instance, metal-on-metal films are used in electronics, data storage and catalysis. In particular, the use of ultrathin magnetic films as data storage devices requires a high degree of accuracy and spatial resolution in the magnetization control. Hence, a lot of experimental and theoretical effort has been devoted in the last years\cite{DeBell} to understand both the equlibrium and out-of-equilibrium properties of these kind of systems. Realistic theoretical descriptions of these systems include both the exchange and the dipolar interactions between the microscopic spins. The competition between long ranged 
antiferromagnetic dipolar interactions and short ranged ferromagnetic exchange 
interactions can give rise to   a variety of unusual and interesting macroscopic  phenomena. Works in two dimensional uniaxial spin systems, where the spins are oriented 
perpendicular to the lattice and coupled with these kind of interactions, 
have shown a very rich phenomenological scenario concerning both its 
equilibrium statistical mechanics\cite{Kashuba,MacIsaac} and non-equilibrium 
dynamical properties\cite{Sampaio,Toloza,Stariolo}.  In particular, some of these results\cite{Toloza} showed the existence of different types of slow relaxation dynamics when the system is quenched from a high-temperature configuration to a subcritical temperature, depending on the relative strenghts of the dipolar and exchange interactions. This change in the relaxation properties could be an effect of a change in the low temperature equilibrium properties of the system, such as the presence of metastable states. In this work we investigate in detail the low temperature phase diagram of that system in the region where the change in the relaxation properties has  been observed.

The above mentioned system is described by the Ising like Hamiltonian  

\begin{equation}
\label{hamilton}
H = - \delta\sum_{<i,j>}{\sigma_i\sigma_j} + 
\sum_{(i,j)}{\frac{\sigma_i\sigma_j}{r_{ij}^3}}
\end{equation}
where the spin variable $\sigma_i=\pm 1$ is located at the site $i$ 
of a square lattice, the sum $\sum_{<i,j>}$ runs over all pairs of nearest 
neighbor sites and the sum $\sum_{(i,j)}$ runs over all distinct pair of 
sites of the lattice; $r_{ij}$ is the distance (in crystal units) between 
sites $i$ and $j$; $\delta$ represents the quotient between the exchange $J_0$ and dipolar $J_d$  coupling parameters, where the energy is measured in units of $J_d$, which is always assumed  to be antiferromagnetic ($J_d>0$). Hence, $\delta>0$ means ferromagnetic exchange coupling. 

 The overall features of the  the finite temperature phase 
diagram of this model were described by MacIsaac and coauthors\cite{MacIsaac} by means of Monte Carlo 
 calculations  on $16\times 16$ lattices  and analytic calculations of the ground state\cite{nota}.

\noindent They found  that the ground state  of Hamiltonian (\ref{hamilton}) is the 
antiferromagnetic state for $\delta<0.425$. For $\delta>0.425$ the 
antiferromagnetic state becomes unstable with respect to the formation of 
striped domain structures, that is, to state configurations with spins 
aligned along a particular axis forming a ferromagnetic stripe of constant 
width $h$, so that spins in adjacent  stripes are anti-aligned,  forming a 
super lattice in the direction perpendicular to the stripes. They also showed that striped states of increasingly higher thickness $h$ become more stable as $\delta$ increases from $\delta=0.425$. Moreover, they showed that the striped states are also more stable than the ferromagnetic one for arbitrary large values of $\delta$, suggesting such a phase to be the ground state of the model for $\delta>0.425$. Monte Carlo calculations on finite lattices at low temperature\cite{MacIsaac,Sampaio} gave further support to this proposal, at least for intermediate values of $\delta$. Furthermore, such simulations have shown that  striped phases of increasingly higher values of $h$ may become thermodynamically stable at  finite temperatures for intermediate values of $\delta$. These results are in agreement with other analytic ones\cite{Kashuba,Chayes}. For low values of $\delta$ the system presents an antiferromagnetic phase at low temperatures. At high temperatures, of course, the system always becomes paramagnetic. Specific heat calculations showed  that the transition between the paramagnetic and the striped phases is a second order one\cite{MacIsaac}, while the nature of the transitions between the different striped phases was not clearly determined.

We performed Monte Carlo simulations of Hamiltonian (\ref{hamilton}) on square lattices up to $48\times 48$ sites using periodic boundary conditions and heat bath dynamics. Our calculations focused on the low temperature region of the $(\delta,T)$ phase diagram for values of $\delta$ between $0.4$ and $2$, which includes the transition line between the $h=1$ (h1) and the $h=2$ (h2)  striped phases.

\begin{center}
\begin{figure}
\epsfig{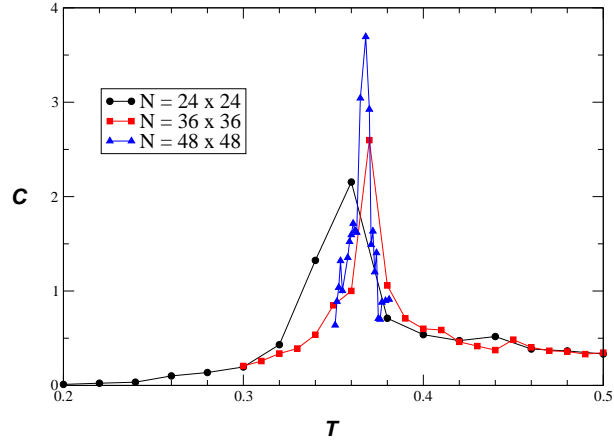}
\caption{Specific heat $C$ {\it vs.} temperature $T$ for $\delta=1.1$ and different system sizes $N$.}
\label{fig1}
\end{figure}
\end{center}

First  we calculated through the energy fluctuations the specific heat $C$ as a 
function of the temperature for different values of $\delta$ and different
system sizes up to $48\times 48$ sites. The typical behaviour of $C$ is shown
in Fig.\ref{fig1} for $\delta=1.1$. By considering the peaks in the specific
heat we obtained the second order critical line between the paramagnetic and
the low temperature ordered phases (h1 and h2). These results (triangles in
Fig.\ref{fig4}) slightly improve  those obtained by MacIsaac and
coauthors\cite{MacIsaac} for $16\times 16$ lattices, thus showing a fast
convergence of the critical temperature for increasing sistem sizes, at least
for small values of $\delta$.
 
\vspace{0.5cm}

Next, we analized the transition between the h1 and h2 phases. To do this we introduced  
the staggered magnetization

\begin{equation}
m_{h1} = \frac{1}{N} \sum_{x,y=1}^N (-1)^x  \sigma_{xy},  
\end{equation}

\noindent where $N$ is the total number of lattice sites; $m_{h1}$ equals one in the ground state corresponding to 
 the striped phase with  $h=1$. We calculated the thermal average
$M_{h1}\equiv <m_{h1}>$ as a function of $T$ as well as the fluctuations of
this quantity through  the associated staggered susceptibility

\begin{equation}
\chi_{h1} = \left< m_{h1}^2 \right> - \left< m_{h1} \right>^2.
\end{equation}

\begin{center}
\begin{figure}
\epsfig{file=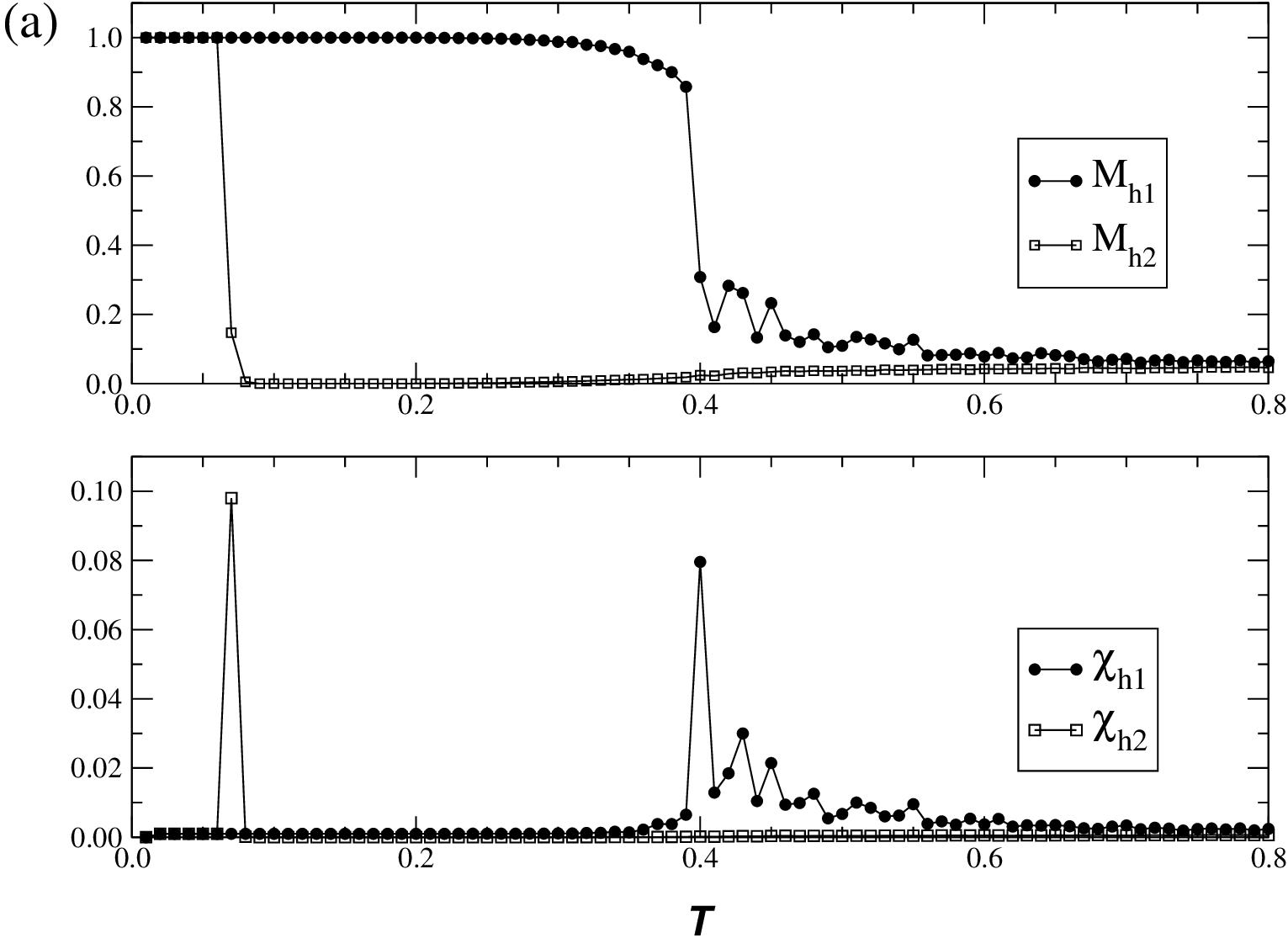,width=8cm} \hspace{0.5cm}
\epsfig{file=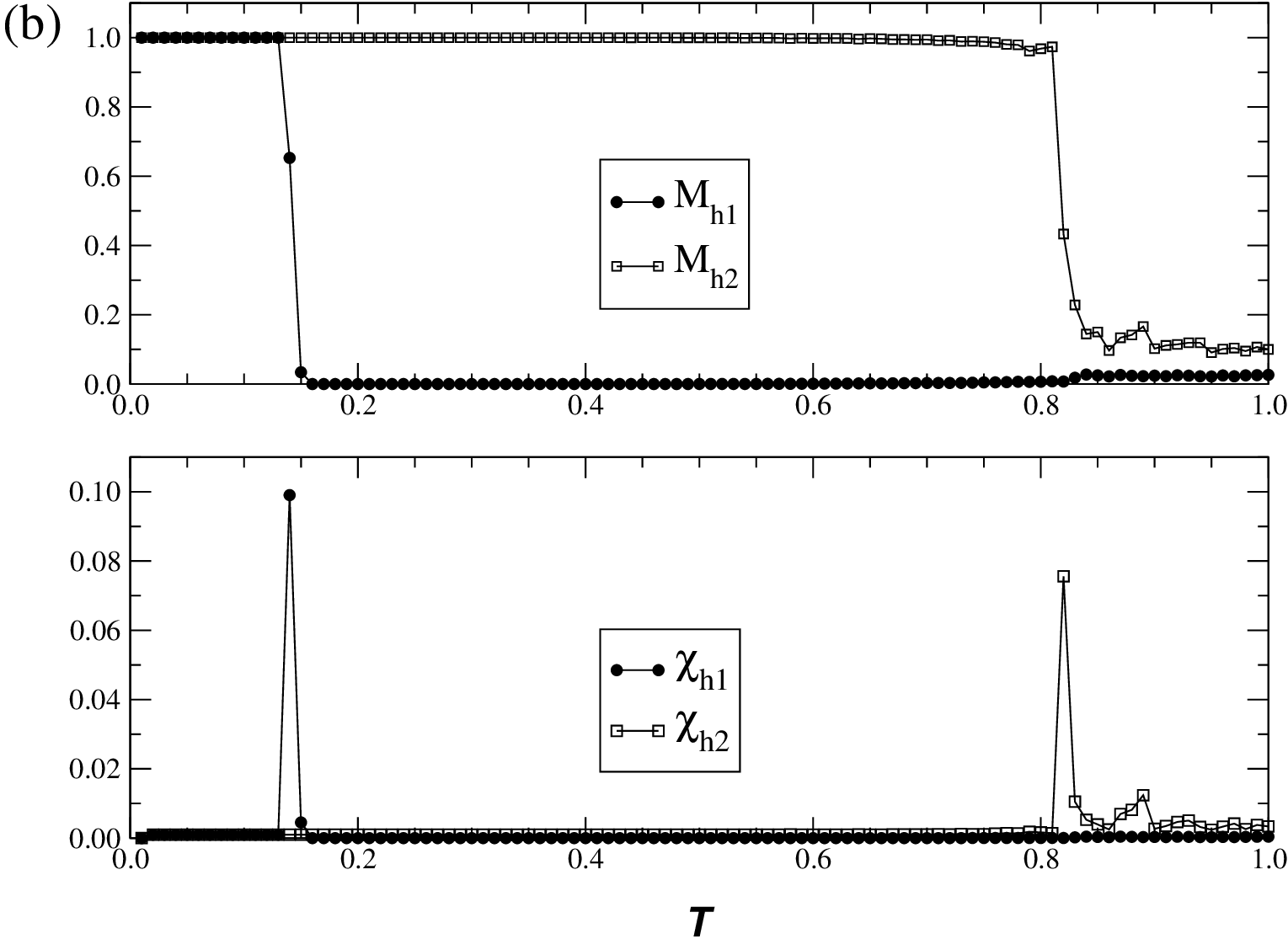,width=8cm}
\caption{Staggered magnetization $M_{h1}$ ($M_{h2}$) and associated susceptibility $\chi_{h1}$ ($\chi_{h2}$) {\it vs.} $T$ when the system is heated starting at $T=0$ from an initial  configuration in the corresponding ground state h1 (h2), for $N=24\times 24$. (a) $\delta=1$ corresponding to the stable phase h1; (b) $\delta=2$ corresponding to the stable phase h2.}
\label{fig2}
\end{figure}
\end{center}

\noindent In a similar way we defined the staggered magnetization $M_{h2}$ and its 
associated susceptibility $\chi_{h2}$, corresponding to the striped phase h2.
We simulated the heating of the system from $T=0$  to a temperature higher
than the critical one for different values of $\delta$ and starting from
different initial configurations. To analize the stability of both phases in
the different parts of the phase diagram we first calculated the staggered
magnetization $M_{h1}$ and susceptibility $\chi_{h1}$ starting from a striped
configuration h1; we also calculated the staggered magnetization
$M_{h2}$ and susceptibility $\chi_{h2}$ starting in a striped configuration
h2. The typical behaviours are shown in Fig.2 for $\delta=1$ (Fig.2a), where
the ground state corresponds to h1, and for $\delta=2$ (Fig.2b), where the
ground state corresponds to h2. For $\delta=1$ we see that $M_{h1}$ falls down
to zero at the critical temperature while $\chi_{h1}$ shows a sharp peak at
the same temperature, as expected for the order parameter and its conjugated
susceptibility  at this type of transition. On the other hand, $M_{h2}$ remains stable
up to some temperature $T_2<T_c$, where it losses stability and falls down
abrubtly to zero, while $\chi_{h2}$ is zero for almost every temperature,
except near $T=T_2$. A similar behavior is observed for $\delta=2$, but with
$(M_{h1},\chi_{h1})$ and $(M_{h2},\chi_{h2})$ interchanged. In this case the
phase h1 losses stability at some temperature $T_1<T_c(\delta)$.

 All these results show the existence at low temperatures of
metastable states h2 in the region of the phase diagram corresponding to the
stable phase h1 and {\it viceversa}. The numerical calculation of  stability
lines $T_1(\delta)$ and $T_2(\delta)$  is depicted in Fig.\ref{fig4} by means
of circles and squares respectively. We see that these lines cross at a certain
value $\delta \approx 1.26$ where they join smoothly with the second order
critical lines.

All these results suggest the existence of a first order phase transition line separating the phases h1 and h2. To verify this we considered the free energy

\begin{equation}
F = U - T \int_0^T \frac{C(T')}{T'} dT'
\end{equation}

\begin{center}
\begin{figure}
\epsfig{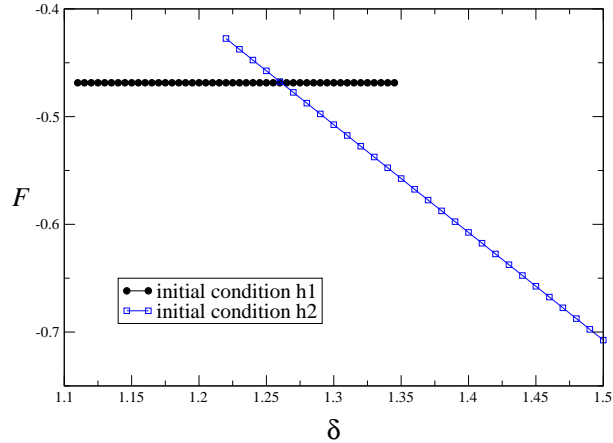}
\caption{Free energy $F$ of the phases h1 and h2 {\it vs.} $\delta$ for $T=0.2$.}
\label{fig3}
\end{figure}
\end{center}

\noindent where $U=\left< H\right> $. We then calculated the free energy of both phases 
h1 and h2 by heating the system from $T=0$ up to a fixed value of temperature
$T$, starting from two different initial configurations corresponding to the
ground state of h1 and h2, and for different values of $\delta$. In
Fig.\ref{fig3} we see an example of this calculation for $T=0.2$. The free
energy of the phase h1 was calculated for increasing values of $\delta$,
starting from some small value well inside the region where this phase is
stable, up to the the value of $\delta$ corresponding to the stability line of
h1 at the given temperature. The same calculation was repeated a the same
temperature for the free energy of the phase h2, but for values of $\delta$
ranging from the stability line up to some value well inside the region where
h2 is stable. The observed continuous change of the minimal free energy from
one phase to the other, with a discontinuous change in the slope as $\delta$ is
varied is a clear evidence of a first order phase transition. Also the
multivalued nature of the free energy gives further evidence of the metastable
nature of these phases in some parts of the phase diagram. Repeating these
calculations for different values of $T$ we obtained the almost vertical first
order transition line between the two phases, showed  in Fig.\ref{fig4} by means of
diamonds. The shaded region in Fig.\ref{fig4} indicates the presence of metastable
states.

\begin{center}
\begin{figure}
\epsfig{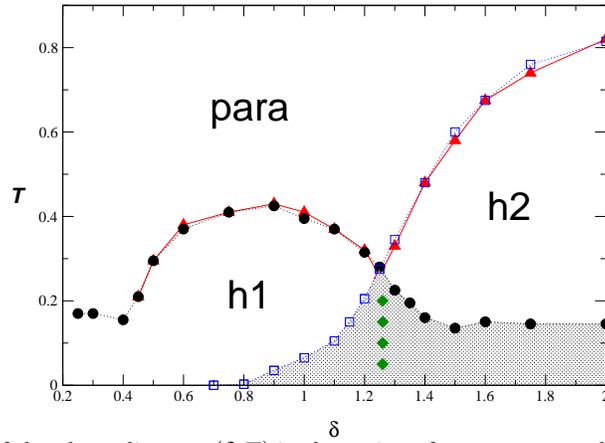}
\caption{Numerical calculation of the phase diagram $(\delta,T)$ in the region of parameters under study. Filled triangles correspond to the critical temperatures $T_c(\delta)$ obtained by specific heat calculations for the phase transition between the ordered striped phases h1 and h2 and the paramagnetic one. Filled circles (open squares) correspond to the stability line of the h1 (h2) phase, obtained by analizing the staggered magnetization $M_{h1}$ ($M_{h2}$). Filled diamonds correspond to the first order transition line between the h1 and h2 phases, obtained by free energy numerical calculations. The shaded region correspond to the presence of metastable states.}
\label{fig4}
\end{figure}
\end{center}

\begin{center}
\begin{figure}
\epsfig{file=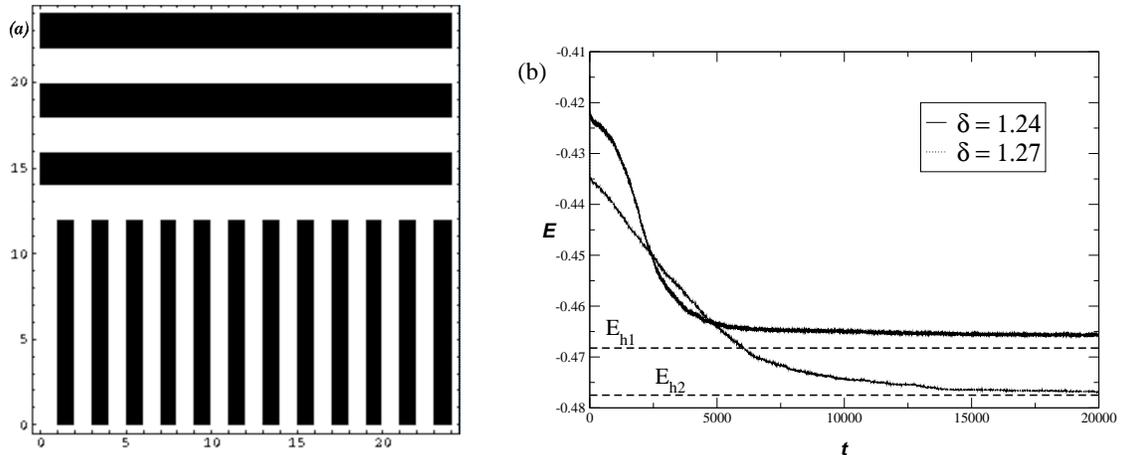,width=6cm} \hspace{0.5cm}
\epsfig{file=fig5b.eps,width=8cm}
\caption{(a) Initial configuration with 50\% in the ground state corresponding to h1 and 50\% in the ground state corresponding to h2. (b) Time evolution of the energy of the system $E$ starting from the initial configuration of figure (a), for $T=0.2$ and $N=24 \times 24$.}
\label{fig5}
\end{figure}
\end{center}

Finally, we considered the relaxation of the system in the metastable region starting 
from a non-uniform initial configuration. We prepared the system in a
configuration with one half of the system in the h1 phase and the other half
in the h2 phase, as shown in Fig.\ref{fig5}. This particular configuration
facilitates the nucleation of the stable phase whatever it be (h1 or h2). In
the same figure  we show also the time evolution of the energy for two
different values of $\delta$ located at both sides of the transition line
$\delta=1.26$. We see that the energy evolves towards the mean energy of h1
($E_{h1}$) at the corresponding temperature when $\delta<1.26$, while it 
relaxes to the mean energy of h2 ($E_{h2}$) when $\delta>1.26$. This shows
that indeed the metastable configurations are unstable agains the nucleation
of the corresponding stable phase.

Summarizing, we presented numerical evidence of the existence of metastable states in 
the model described by Hamiltonian (\ref{hamilton})   in the low
temperature region of the $(\delta,T)$ phase diagram where the transition
between h1 and h2 phases takes place. We also showed that these metastable
states are associated with a first order phase transition line between both
phases. The presence of these metastable states may alter the normal domain
growth process  when the system is quenched from high temperatures into the 
ordered phase, by pinning the walls of the stable phase domains. Such process
could slow down the normal coarsening dynamics, depending on whether the quench
drops the system into the metastable region or not,  thus explaining the
observed  change in the relaxation dynamics in this region of the phase
space\cite{Toloza}. Some work along this line is in progress and will be
published elsewhere. Another point that deserves further investigation is the
possible presence of more complex metastable states for higher values of
$\delta$, as observed in a related three-dimensional model with competing
nearest neighbors ferromagnetic interactions and long-range antiferromagnetic
Coulomb-like interactions\cite{Grousson}.

  This work was partially
supported by grants from
 Consejo Nacional de Investigaciones Cient\'\i ficas
y T\'ecnicas CONICET 
 (Argentina),  Agencia C\'ordoba Ciencia (C\'ordoba,
Argentina) and  Secretar\'\i a de Ciencia y 
 Tecnolog\'\i a de la Universidad
Nacional de C\'ordoba (Argentina).


\begin{references}
\bibitem{nota} It is worth noting
that MacIsaac and coauthors\cite{MacIsaac} definition of the Hamiltonian
(Eq.(5) in that reference) is slightlly different from ours
(Eq.(\ref{hamilton})).While in that paper the dipolar term contains a sum over
all pairs of spins, in Hamiltonian (\ref{hamilton}) we consider the sum over
every pair of spins just once. This leads to the equivalence $\delta=J/2$, $J$
being the echange parameter in the above reference. Since the dipolar
parameter also fixes the energy units in our work, there is also a factor
$1/2$ between the critical temperatures obtained in both works.
\end{references}
\end{document}